\documentclass[aps,twocolumn,preprintnumbers,showpacs]{revtex4}
\usepackage{mathrsfs}
\usepackage{longtable,lscape}
\usepackage{txfonts}
\usepackage{amssymb}
\usepackage{indentfirst}
\usepackage{graphicx,,booktabs}
\usepackage{color}
\usepackage{amssymb}
\usepackage{epsfig}
\newcommand{\vsig}{\mbox{\boldmath$\sigma$\unboldmath}}
\newcommand{\veps}{\mbox{\boldmath$\epsilon$\unboldmath}}

\begin{document}

\title{$\eta$ photoproduction on the quasi-free nucleons in the chiral quark model}
\author{
Xian-Hui Zhong$^{1}$ \footnote {E-mail: zhongxh@ihep.ac.cn} and
Qiang Zhao$^{2,3}$ \footnote {E-mail: zhaoq@ihep.ac.cn}}

\affiliation{ 1)  Department of Physics, Hunan Normal University,
and Key Laboratory of Low-Dimensional Quantum Structures and Quantum
Control of Ministry of Education, Changsha 410081, P.R. China }

\affiliation{ 2) Institute of High Energy Physics,
       Chinese Academy of Sciences, Beijing 100049, P.R. China
}
\affiliation{3) Theoretical Physics Center for Science Facilities,
Chinese Academy of Sciences, Beijing 100049, P.R. China}


\begin{abstract}

A chiral quark-model approach is adopted to study the $\eta$
photoproduction off the quasi-free neutron and proton from a
deuteron target. Good descriptions of the differential cross
sections, total cross sections and beam asymmetries for these two
processes are obtained in the low energy region. For $\gamma
p\rightarrow \eta p$, the dominant resonances are $S_{11}(1535)$,
$S_{11}(1650)$, $D_{13}(1520)$, $D_{13}(1700)$ and $P_{13}(1720)$.
For the $\gamma n\rightarrow \eta n$ process, the dominant
resonances are $S_{11}(1535)$, $S_{11}(1650)$, $D_{13}(1520)$,
$D_{15}(1675)$ and $P_{13}(1720)$. Furthermore, the $u$ channel
backgrounds have significant contributions to the $\eta$
photoproduction processes. The configuration mixings in
$S_{11}(1535,1650)$ and $D_{13}(1520,1700)$ can be extracted, i.e.
$\theta_S\simeq 26^\circ$ and $\theta_D\simeq 21^\circ$. It shows
that the narrow bump-like structure around $W= 1.68$ GeV observed in
$\gamma n\rightarrow \eta n$ can be naturally explained by the
constructive interferences between $S_{11}(1535)$ and
$S_{11}(1650)$. In contrast, the destructive interference between
$S_{11}(1535)$ and $S_{11}(1650)$ produces the shallow dip around
$W= 1.67$ GeV in $\gamma p\rightarrow \eta p$. The $S$ wave
interfering behaviors in the proton and neutron reactions are
correlated with each other in the quark model framework, and no new
exotic nucleon resonances are needed in these two reactions. The
helicity amplitudes of $S_{11}(1535)$, $S_{11}(1650)$,
$D_{13}(1520)$, $D_{13}(1700)$ and $D_{15}(1675)$ are extracted from
the reactions as well.

\end{abstract}


\maketitle

\section{Introduction}

Understanding of the baryon spectrum and the search for the missing
nucleon resonances and new exotic states are hot topics in nuclear
physics~\cite{Klempt:2009pi}. Photoproduction of mesons is an ideal
tool for the study of nucleon resonances in
experiment~\cite{Krusche:2003ik}. Recently, a bump structure around
$W=1.68$ GeV was observed in the excitation function of $\eta$
production off quasi-free neutrons by Ref.~\cite{Kuznetsov:2004gy},
which was considered as an evidence for a narrow resonance. To
further understand the ``anomalous" peak around $W=1.68$ GeV, the
measurement of the polarized beam asymmetry in $\eta$
photoproduction on the neutron is also carried out by the GRAAL
Collaboration~\cite{Fantini:2008zz}. The bump-like structure
observed in $\gamma n\rightarrow \eta n$ stimulates great interests
of other experimental collaborations. The CBELSA/TAPS measured the
quasi-free photoproduction of $\eta$ mesons off nucleons bound in
the deuteron~\cite{Jaegle:2008ux}, and found that there was a
pronounced bump-like structure at $W\simeq 1.68$ GeV in the
excitation function for $\eta$ production off the neutron. This
structure was absent in $\gamma p\rightarrow \eta p$. This structure
was also confirmed in the $\eta$ production off the neutron by the
LNS \cite{Miyahara:2007zz} and Crystal
Ball/TAPS~\cite{Werthmuller:2010af}. Very recently, the quasi-free
compton scattering on the neutron was measured. The data reveals a
narrow peak at $W\simeq 1.68$ GeV as well, which is considered as
evidence for a narrow structure in the $\gamma n\rightarrow \eta n$
in association with the photoproduction
data~\cite{Kuznetsov:2010as}.

Explanations for the bump structure around $W\simeq 1.68$ GeV have
been proposed in the literature. One
proposal~\cite{Azimov:2005jj,Fix:2007st,Choi:2005ki,Kuznetsov:2007gr,Kuznetsov:2011pe}
is that the narrow structure might correspond to the antidecuplet of
pentaquark states. The other possibility is that the narrow
structure around $W\simeq 1.68$ GeV is due to the interferences of
some well known nucleon resonances. For example, Klempt \emph{et
al.} predict that the narrow structure can be naturally interpreted
by interferences between $S_{11}(1535)$ and
$S_{11}(1650)$~\cite{Anisovich:2008wd}. Similar conclusions are
obtained by Shyam and Scholten within a coupled-channels K-matrix
approach, where the the bump structure at $W\simeq 1.68$ GeV is
attributed to the interfering effects of $S_{11}(1535)$,
$S_{11}(1650)$, $P_{11}(1710)$ and
$P_{13}(1710)$~\cite{Shyam:2008fr}. In contrast, Shklyar \emph{et
al.} suggest that the narrow structure can be explained in terms of
coupled-channel effects due to $S_{11}(1650)$ and $P_{11}(1710)$
resonance excitations~\cite{Shklyar:2006xw}.

It is also interesting to ask why the bump structure has not been
observed in $\gamma p\rightarrow \eta p$ if it indeed corresponds to
a genuine resonance. Kuznetsov \emph{et al.} presented their new
data of the beam asymmetry for $\eta$ photoproduction on free
protons, which may revealed a structure~\cite{Kuznetsov:2007dy}.
However, it is questionable to explain this structure as an evidence
of a narrow resonance according to the partial wave
analysis~\cite{Anisovich:2008wd}. Very recently the MAMI-C
Collaboration reported their new observations on $\gamma
p\rightarrow \eta p$. A shallow dip near $W=1.68$ GeV in the total
cross section is found to be caused by a substantial dip in $\eta$
production at forward angles~\cite{McNicoll:2010qk}. It is wondered
whether the dip structure is in connection with the observed narrow
structure in the neutron reaction. To have a better understanding of
these questions, a combined study of the $\eta$ production off
quasi-free protons and neutrons is urgently needed.

In this work, we present a systemic analysis of experimental
observables, such as differential cross sections, total cross
sections, and polarized beam asymmetry, for the $\eta$
photoproduction on the quasi-free nucleons, within a constituent
quark model.

In the quark model, an effective chiral Lagrangian is introduced to
account for the quark-pseudoscalar-meson coupling. Since the
quark-meson coupling is invariant under the chiral transformation,
some of the low-energy properties of QCD are retained. The chiral
quark model has been well developed and widely applied to meson
photoproduction reactions
~\cite{qk1,qk2,qkk,Li:1997gda,qkk2,qk3,qk4,qk5,Li:1995vi,Li:1998ni,
Saghai:2001yd,He:2008ty,He:2009zzi}. It has several obvious
features. One is that only a limited number of parameters will
appear in the formalism. In principle, only one parameter is needed
for the resonances to be coupled to the pseudoscalar meson in the
SU(6)$\otimes$O(3) symmetric quark model limit. This distinguishes
it from hadronic models where each resonance requires one individual
coupling constant as a free parameter. Secondly, since all the
resonances are treated equivalently at the quark level. Thus, it
would have predictive powers to some extent when exposed to
experimental data. Information about the resonance structures and
form factors can thus be extracted. Meanwhile, insights into the
SU(6)$\otimes$O(3) symmetry breaking can be gained.

In the case of $\eta$ meson photoproduction off nucleons, the chiral
quark model can be applied and transition amplitudes at the tree
level can be explicitly calculated. There are interesting
differences between  $\gamma p\rightarrow \eta p$ and  $\gamma
n\rightarrow \eta n$. In the $\gamma p$ reactions, contributions
from states of representation $[70,^48]$ will be forbidden by the
Moorhouse selection rule~\cite{Moorhouse:1966jn}. As a consequence,
only states of $[56,^28]$ and $[70,^28]$ would contribute to the
$\eta$ production. In contrast, all the octet states can contribute
to the $\gamma n$ reactions. In another word, more states will be
present in the $\gamma n$ reactions. Therefore, by studying the
$\eta$ meson photoproduction on the quasi-free nucleons, we expect
that the role played by intermediate baryon resonances can be
highlighted.

The paper is organized as follows. In Sec.~\ref{fram}, a brief
review of the chiral quark model approach is given. The numerical
results are presented and discussed in Sec.~\ref{cy}. Finally, a
summary is given in Sec.~\ref{summ}.

\section{framework}\label{fram}

In the chiral quark model, the quark-pseudoscalar-meson and
electromagnetic couplings at the tree level are described by
\cite{Li:1997gda,qk3}
\begin{eqnarray}\label{lagrangian-1}
H_m&=&\sum_j
\frac{1}{f_m}\bar{\psi}_j\gamma^{j}_{\mu}\gamma^{j}_{5}\psi_j\vec{\tau}
\cdot\partial^{\mu}\vec{\phi}_m,\label{coup}\\
H_e&=&-\sum_je_j\gamma^{j}_{\mu}A^{\mu}(\mathbf{k},\mathbf{r})\label{coup1},
\end{eqnarray}
where $\psi_j$ represents the $j$-th quark field in a hadron, and
$f_m$ is the meson's decay constant. The pseudoscalar-meson octet
$\phi_m$ is written as
\begin{eqnarray}
\phi_m=\pmatrix{
 \frac{1}{\sqrt{2}}\pi^0+\frac{1}{\sqrt{6}}\eta & \pi^+ & K^+ \cr
 \pi^- & -\frac{1}{\sqrt{2}}\pi^0+\frac{1}{\sqrt{6}}\eta & K^0 \cr
 K^- & \bar{K}^0 & -\sqrt{\frac{2}{3}}\eta}.
\end{eqnarray}

The $s$ and $u$ channel transition amplitudes are determined by
\begin{eqnarray}
\mathcal{M}_{s}=\sum_j\langle N_f |H_{m} |N_j\rangle\langle N_j
|\frac{1}{E_i+\omega_\gamma-E_j}H_{e}|N_i\rangle,\\
\mathcal{M}_{u}=\sum_j\langle N_f |H_{e }
\frac{1}{E_i-\omega_m-E_j}|N_j\rangle\langle N_j | H_{m }
|N_i\rangle,
\end{eqnarray}
where $\omega_\gamma$ is the energy of the incoming photons.
$|N_i\rangle$, $|N_j\rangle$ and $|N_f\rangle$ stand for the
initial, intermediate and final states, respectively, and their
corresponding energies are $E_i$, $E_j$ and $E_f$, which are the
eigenvalues of the NRCQM Hamiltonian $\hat{H}$~\cite{Isgur:1978xj,
Isgur:1977ef}.  The $s$ and $u$ channel transition amplitudes have
been worked out in the harmonic oscillator basis in Ref.
\cite{Li:1997gda}. The $t$ channel contributions due to vector meson
exchange are not considered in this work. If a complete set of
resonances are included in the $s$ and $u$ channels, the
introduction of $t$ channel contributions might result in double
counting~\cite{Dolen:1967jr}. In fact, the $t$ channel vector-meson
exchange contribution was found to be negligible at low energy
region~\cite{qk3,He:2007mw}.

In the Chew-Goldberger-Low-Nambu (CGLN)
parameterization~\cite{Chew:1957tf}, the transition amplitude can be
written with a standard form:
\begin{eqnarray}
\mathcal{M}&=&i f_1 \vsig \cdot \veps+f_2 \frac{(\vsig \cdot
\mathbf{q})\vsig\cdot (\mathbf{k}\times
\veps)}{|\mathbf{q}||\mathbf{k}|}\nonumber\\
&& +if_3\frac{(\vsig \cdot \mathbf{k}) (\mathbf{q}\cdot
\veps)}{|\mathbf{q}||\mathbf{k}|}+if_4\frac{(\vsig \cdot \mathbf{q})
(\mathbf{q}\cdot \veps)}{|\mathbf{q}|^2},
\end{eqnarray}
where $\vsig$ is the spin operator of the nucleon, $\veps$ is the
polarization vector of the photon, and $\mathbf{k}$ and $\mathbf{q}$
are incoming photon and outgoing meson momenta, respectively.

It is should be remarked that the amplitudes in terms of the
harmonic oscillator principle quantum number $n$ are the sum of a
set of SU(6) multiplets with the same $n$. To see the contributions
of individual resonances, we need to separate out the
single-resonance-excitation amplitudes within each principle number
$n$ in the $s$-channel. Taking into account the width effects of the
resonances, the resonance transition amplitudes of the $s$-channel
can be generally expressed as \cite{Li:1997gda}
\begin{eqnarray}
\mathcal{M}^s_R=\frac{2M_R}{s-M^2_R+iM_R
\Gamma_R(\mathbf{q})}\mathcal{O}_Re^{-(\textbf{k}^2+\textbf{q}^2)/6\alpha^2},
\label{stt}
\end{eqnarray}
where $\sqrt{s}=E_i+\omega_\gamma$ is the total energy of the
system, $\alpha$ is the harmonic oscillator strength, $M_R$ is the
mass of the $s$-channel resonance with a width
$\Gamma_R(\mathbf{q})$, and $\mathcal{O}_R$ is the separated
operators for individual resonances in the $s$-channel. Its general
structure is given by \cite{Li:1997gda},
\begin{eqnarray}
\mathcal{O}_R&=&g_R A[f_1^R\vsig \cdot \veps+if_2^R(\vsig \cdot
\mathbf{q})\vsig\cdot (\mathbf{k}\times \veps)\nonumber\\
&&+f_3^R(\vsig \cdot \mathbf{k}) (\mathbf{q}\cdot \veps)+f_4^R(\vsig
\cdot \mathbf{q}) (\mathbf{q}\cdot \veps)],
\end{eqnarray}
where $g_R$ is an isospin factor, $A$ is the meson decay amplitude,
and $f_i^R(i=1,2,3,4)$ is proportional to the photon transition
amplitude. The detail of extracting $\mathcal{O}_R$ can be found in
\cite{Li:1997gda}.

Finally, the physical observables, differential cross section and
photon beam asymmetry, are given by the following standard
expressions~\cite{Walker:1968xu}:
\begin{eqnarray}
\frac{d\sigma}{d\Omega}&=&\frac{\alpha_e\alpha_m(E_i+M_N)(E_f+M_N)}{16s
M_N^2}\frac{1}{2}\frac{|\mathbf{q}|}{|\mathbf{k}|}\sum^4_i|H_i|^2,\label{edf}\\
\Sigma &=&2\mathrm{Re}(H_4^*H_1-H_3^*H_2)/\sum^4_i|H_i|^2,
\end{eqnarray}
where $\alpha_m$ is the meson-nucleon-nucleon coupling constant, and
$\alpha_e$ is the fine-structure constant. The helicity amplitudes
$H_i$ can be expressed by the CGLN amplitudes $f_i$ with
relations~\cite{Walker:1968xu}:
\begin{eqnarray}
H_1&=&-\frac{1}{\sqrt{2}}\sin\theta\cos\frac{\theta}{2}(f_3+f_4),\\
H_2&=&\sqrt{2}\cos\frac{\theta}{2}[(f_2-f_1)+\sin^2\frac{\theta}{2}(f_3-f_4)],\\
H_3&=&\frac{1}{\sqrt{2}}\sin\theta\sin\frac{\theta}{2}(f_3-f_4),\\
H_4&=&\sqrt{2}\sin\frac{\theta}{2}[(f_2+f_1)+\cos^2\frac{\theta}{2}(f_3-f_4)].
\end{eqnarray}

In this work, we study the $\eta$ mesons photoproduction off
quasi-free proton and neutron from a deuteron target. The deuteron
is at rest in the laboratory system, but the nucleons are not. Thus,
the Fermi motion of the quasi-free nucleon should be included. To
take into account the Fermi motion, we follow the method of
Ref.~\cite{Anisovich:2008wd} and fold the cross section for the free
nucleon case with the momentum distribution of the nucleon inside
the deuteron. This method is also adopted in
Ref.~\cite{Doring:2009qr}.

The quasi-free nucleon bound in the deuteron has a internal Fermi
momentum $\mathbf{p}_N$. Its three components are defined as
\begin{eqnarray}
p_{xN}&=&|\mathbf{p}_N|\sin\Theta_N\cos\phi_N,\nonumber\\
p_{yN}&=&|\mathbf{p}_N|\sin\Theta_N\sin\phi_N,\nonumber\\
p_{zN}&=&|\mathbf{p}_N|z_N.
\end{eqnarray}
The outgoing meson momentum $\mathbf{q}$ has a relation with the
internal Fermi momentum $\mathbf{p}_N$~\cite{Anisovich:2008wd}:
\begin{eqnarray}
|\mathbf{q}|=\frac{\Sigma\xi|\mathbf{P}^m|+P^m_0\sqrt{\Sigma^2-m_1^2[(P^m_0)^2-
|\mathbf{P}^m|^2\xi]}}{(P^m_0)^2-|\mathbf{P}^m|^2\xi},
\end{eqnarray}
where we have defined
\begin{eqnarray}
\Sigma&=&\frac{1}{2}(s_{tot}-m_1^2-M_N^2),\\
\xi&=&\frac{zP^m_z+|\mathbf{p}_N|\sqrt{1-z_N^2}\sqrt{1-z^2}\cos\phi_N}{|\mathbf{P}^m|},\\
|\mathbf{P}^m|&=&\frac{M_N(p_{zN}+\omega_\gamma)-\omega_\gamma(p_{0N}-p_{zN})}{\sqrt{s_{eff}}}+\omega_\gamma,\\
P_0^m&=&\frac{M_N(p_{0N}+\omega_\gamma)+\omega_\gamma(p_{0N}-p_{zN})}{\sqrt{s_{eff}}}.
\end{eqnarray}
In the above equations, $z$ is the cosine of the angle between meson
and photon, i.e. $z\equiv \cos\theta$, $m_1$ is the mass of the
outgoing meson, and several variables are defined as
\begin{eqnarray}
p_{0N}&=&\sqrt{M_N^2+|\mathbf{p}_N|^2},\\
s_{tot}&=&M_N^2+2\omega_\gamma(p_{0N}-|\mathbf{p}_N|z_N),\\
\sqrt{s_{eff}}&=&\sqrt{M_N^2+2M_N\omega_\gamma},\\
P^m_z&=&
\frac{M_Np_{0N}+\omega_\gamma(p_{0N}-p_{zN}+M_N)}{\sqrt{s_{eff}}}\nonumber\\
          &&-\frac{\sqrt{s_{eff}}}{M_N}(p_{0N}-p_{zN}).
\end{eqnarray}

Thus, the differential cross section for the meson photoproduction
off nucleons bound in a deuteron can be expressed
as~\cite{Anisovich:2008wd}:
\begin{eqnarray}
\frac{d\sigma_{\gamma D}}{d\Omega}=\int \frac{d\sigma}{d\Omega}
f^2(\mathbf{p}_N)|\mathbf{p}_N|^2d|\mathbf{p}_N|\frac{dz_Nd\phi_N}{4\pi},
\end{eqnarray}
where $f^2(\mathbf{p}_N)$ is the nucleon momentum distribution
inside the deuteron. In the calculations, we can choose the simple
parametrization of the Paris ~\cite{Lacombe:1981eg} or CD-Bonn
deuteron functions~\cite{Machleidt:2000ge}. Both of the
parametrizations give nearly the same wave functions for deuteron.

\section{CALCULATIONS AND ANALYSIS} \label{cy}

\subsection{Parameters}

In our framework, the resonance transition operator,
$\mathcal{O}_R$, is derived in the SU(6)$\otimes$O(3) symmetric
quark model limit. In reality, the SU(6)$\otimes$O(3) symmetry is
generally broken due to e.g. spin-dependent forces in the
quark-quark interaction. As a consequence, configuration mixings
would occur, and an analytic solution cannot be achieved. An
empirical way~\cite{Li:1998ni, Saghai:2001yd,He:2008ty,He:2009zzi}
to accommodate the configuration mixings in the symmetric quark
model framework is to introduce a set of coupling strength
parameters, $C_R$, for each resonance amplitude,
\begin{eqnarray}
\mathcal{O}_R\rightarrow C_{R} \mathcal{O}_{R}
 \ ,
\end{eqnarray}
where $C_R$ can be determined by fitting the experimental
observables. In the SU(6)$\otimes$O(3) symmetry limit, one finds
$C_R=1$ while deviations of $C_R$ from unity imply the
SU(6)$\otimes$O(3) symmetry breaking. For those two $S$-wave $1/2^-$
and $D$-wave $3/2^-$ states, state mixings seem to be inevitable. We
explicitly express the transition amplitudes as follows:
\begin{eqnarray}\label{coeff-1}
\mathcal{O}_R\rightarrow C^{[70,^{2}8]}_{R}
\mathcal{O}_{[70,^{2}8,J]}+C^{[70,^{4}8]}_{R}
\mathcal{O}_{[70,^{4}8,J]} \ ,
\end{eqnarray}
where the coefficients $C^{[70,^{2}8]}_{R}$ and $C^{[70,^{4}8]}_{R}$
should contain state mixing information and from which the mixing
angles between $S_{11}(1535)$ and $S_{11}1650)$, and $D_{13}(1520)$
and $D_{13}(1700)$ can be extracted. The determined $C_R$ values for
low-lying resonances are listed in Tab.~\ref{param C}, which are in
agreement with the previous quark model study of
Ref.~\cite{Saghai:2001yd}. We will discuss them in details in the
later subsections.

To take into account the relativistic effects~\cite{qkk}, the
commonly applied Lorentz boost factor is introduced in the resonance
amplitude for the spatial integrals, which is
\begin{eqnarray}
\mathcal{O}_R(\textbf{k},\textbf{q})\rightarrow
\mathcal{O}_R(\gamma_k\textbf{k}, \gamma_q\textbf{q} ),
\end{eqnarray}
where $\gamma_k=M_N/E_i$ and $\gamma_q=M_N/E_f$.

The $\eta NN$ coupling, $\alpha_\eta$, is a free parameter in the
present calculations and to be determined by the experimental data.
In the present work the overall parameter $\eta NN$ coupling
$\alpha_\eta$ is set to be the same for both of the processes
$\gamma n\rightarrow \eta n$ and $\gamma p\rightarrow \eta p$. Our
determined $\eta NN$ coupling, $g_{\eta NN}\simeq 2.13$ (i.e.
$\alpha_\eta\equiv g^2_{\eta NN}/4\pi=0.36$), is in good agreement
with the determinations in
Refs.~\cite{Li:1995vi,Tiator:1994et,Piekarewicz:1993ad,Zhu:2000eh}.

There are another two overall parameters, the constituent quark mass
$m_q$ and the harmonic oscillator strength $\alpha$, from the
transition amplitudes. In the calculation we adopt their standard
values in the the quark model, $m_q=330$ MeV and $\alpha^2=0.16$
GeV$^2$.

In the calculations, we allow the resonance masses and widths to
have some changes around the values from PDG to better describe the
data. The determined values are listed in Tab. \ref{parameter}. It
is found that most of the resonance masses and widths are in the
range of PDG values. The favorable widths of $S_{11}(1535)$ and
$S_{11}(1650)$ are $119$ MeV and 105 MeV, respectively, which are
slightly smaller than the lower limit of the PDG
values~\cite{Amsler:2008zzb}.

\begin{table}[ht]
\caption{The strength parameters $C_R$ determined by the
experimental data.} \label{param C}
\begin{tabular}{|c|c|c|c|c|c| }\hline\hline
parameter    &\ \  $\gamma n\rightarrow \eta n$ \ \ & \ \ $\gamma p\rightarrow \eta p$ &
$\gamma p\rightarrow \eta p$~\cite{Saghai:2001yd} \\
\hline
$C^{[70,^28]}_{S_{11}(1535)}$& 0.85     &1.10    & 1.120     \\
$C^{[70,^48]}_{S_{11}(1535)}$& 1.66     &...     &    \\
$C^{[70,^28]}_{S_{11}(1650)}$& $-0.14$  &$-0.22$ & $-0.200$       \\
$C^{[70,^48]}_{S_{11}(1650)}$& $1.29$   &...     &      \\
$C^{[70,^28]}_{D_{13}(1520)}$& 1.21     &1.80    & 0.964     \\
$C^{[70,^48]}_{D_{13}(1520)}$& 0.78     &...     &   \\
$C^{[70,^28]}_{D_{13}(1700)}$& 0.34     &0.50    & 0.036    \\
$C^{[70,^48]}_{D_{13}(1700)}$& 1.90     &...     &       \\
$C_{D_{15}(1675)}$& 1.80                &...     &                  \\
$C_{P_{13}(1720)}$& 3.00                &0.81     &1.000                 \\
$C_{P_{13}(1900)}$& $-1.00$             &$-1.00$  &$-2.478$                  \\
\hline
\end{tabular}
\end{table}

\begin{table}[ht]
\caption{Breit-Wigner masses $M_R$ (MeV) and widths $\Gamma_R$ (MeV)
for the resonances in the $s$-channel. } \label{parameter}
\begin{tabular}{|c|c|c||c|c|c| }\hline\hline
resonance &\ \  $M_R$ \ \ & \ \ $\Gamma_R$ \ \ &\ \ $M_R$ (PDG)\ \ & \ \ $\Gamma_R$ (PDG)\ \  \\
\hline
$S_{11}(1535)$& 1520    &119   &  $1535\pm 10 $   &$150\pm 25$   \\
$S_{11}(1650)$& 1630    &105   &  $1655^{+15}_{-10} $    & $165\pm 20$  \\
$D_{13}(1520)$& 1525    &100   &  $1520\pm 5$   &$115^{+10}_{-15} $ \\
$D_{13}(1700)$& 1690    &110    &  $1700\pm 50$    &$100\pm 50$ \\
$D_{15}(1675)$& 1675    &140    &  $1675\pm 5$    &$150^{+15}_{-20} $ \\
$P_{13}(1720)$& 1700    &167    &  $1720^{+30}_{-20} $    &$200^{+100}_{-50} $ \\
\hline
\end{tabular}
\end{table}

\begin{figure}[ht]
\centering \epsfxsize=8.8 cm \epsfbox{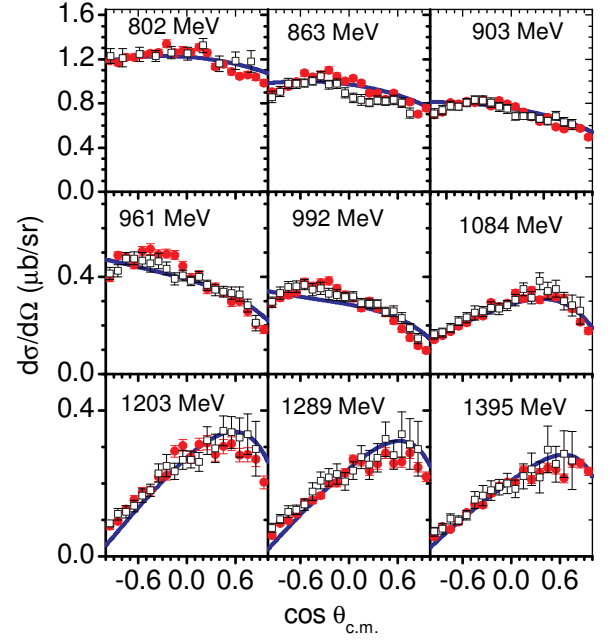} \caption{(Color
online) The differential cross sections for the $\eta$
photoproduction off the free proton at various beam energies. The
data are taken from~\cite{McNicoll:2010qk} (circles) and
~\cite{Bartalini:2007fg} (squares). }\label{fig-5-diff}
\end{figure}

\begin{figure}[ht]
\centering \epsfxsize=8.0 cm \epsfbox{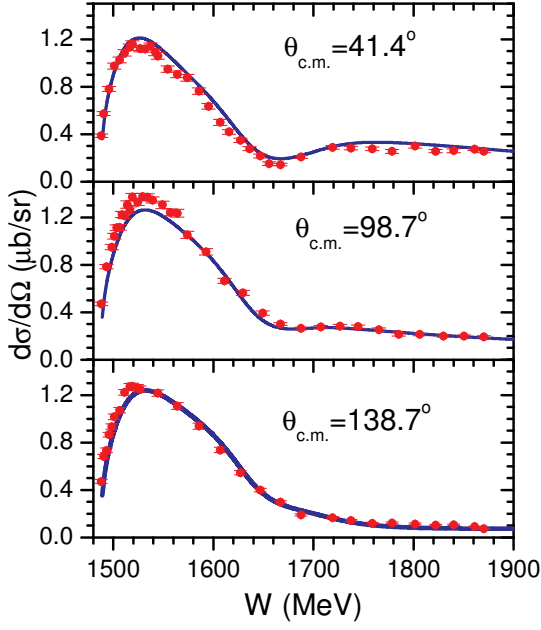} \caption{(Color
online) The fixed-angle excitation functions as a function of center
mass energy $W$ for the $\eta$ photoproduction off the free proton.
The circles stand for the data from~\cite{McNicoll:2010qk}.
}\label{fig-5-diw}
\end{figure}

\begin{figure}[ht]
\centering \epsfxsize=8 cm \epsfbox{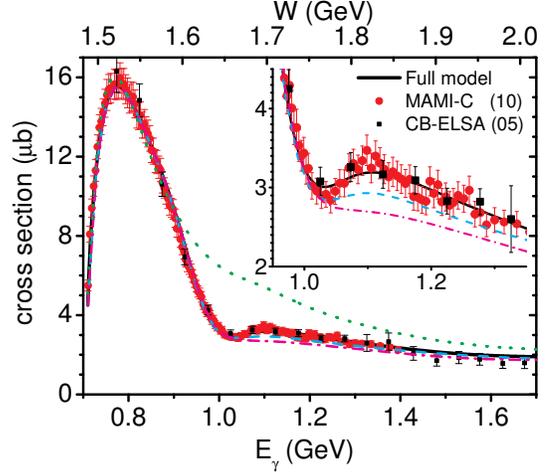} \caption{(Color
online) The total cross section for the $\eta$ photoproduction off
the free proton. The circles and squares stand for the data
from~\cite{McNicoll:2010qk} and ~\cite{Crede:2003ax}, respectively.
The dotted, dashed, dash-dotted curves are for the results by
switching off the contributions from $S_{11}(1650)$, $P_{13}(1720)$
and $D_{13}(1700)$, respectively.}\label{fig-5-tot}
\end{figure}

\subsection{$\gamma p\rightarrow \eta p$ with free proton target}

The chiral quark model studies of $\gamma p\to \eta p$ have been
carried out in
Refs.~\cite{qk4,Li:1995vi,Li:1998ni,Saghai:2001yd,He:2008ty,He:2009zzi}.
The improvement of the experimental situations allows one to
determine the free parameters introduced in the formulation. In this
channel the $N^*$ states of representation $[70, ^48]$ are
prohibited from contribution because of the Moorhouse selection
rule~\cite{Moorhouse:1966jn}. This would significantly reduce the
number of $s$-channel resonances in the reaction in the
SU(6)$\otimes$O(3) symmetry limit. However, the configuration mixing
effects would violate this selection rule, and mixings between
representations of $[70, ^28]$ and $[70, ^48]$ with the same quantum
numbers have to be taken into account in order to obtain a good
description of the data near threshold.

To be more specific, $S_{11}(1535)$ and $S_{11}(1650)$ as mixing
states of $|70,^28, 1/2^-\rangle$ and $|70,^48, 1/2^-\rangle$ can be
expressed as follows:
\begin{eqnarray}\label{mix-s11}
S_{11}(1535)= \cos\theta_S \left|70,^28, 1/2^-\right\rangle-\sin
\theta_S \left|70,^48, 1/2^-\right\rangle ,\nonumber\\
S_{11}(1650)= \sin\theta_S \left|70,^28, 1/2^-\right\rangle + \cos
\theta_S \left|70,^48, 1/2^-\right\rangle ,
\end{eqnarray}
where $\theta_S$ is the mixing angle. Taking into account that the
strong transition ratio $R_S^N\equiv \langle N|H_m|70,^48,
1/2^-\rangle /\langle N|H_m|70,^28, 1/2^-\rangle =-1$, and that the
EM transition amplitude $\langle p |H_e|70,^48 \rangle = 0$ in the
symmetric quark model due to the Moorhouse selection
rule~\cite{Moorhouse:1966jn}, $S_{11}(1535)$ and $S_{11}(1650)$
transition amplitudes will only involve the term of $[70, ^28]$
representations defined in Eq.~(\ref{coeff-1}), of which the
coefficients can be expressed as follows:
\begin{eqnarray}\label{mix-coeff1}
C^{[70,^28]}_{S_{11}(1535)}&=&\cos\theta_S(\cos\theta_S-R_S^N\sin\theta_S),\nonumber\\
C^{[70,^28]}_{S_{11}(1650)}&=&\sin\theta_S(\sin\theta_S+R_S^N\cos\theta_S)
\ .
\end{eqnarray}
By fitting the data for $\gamma p\to \eta p$ to determine the above
coefficients and taking their ratio, one can extract the mixing
angle $\theta_S$. In particular, it shows that the data favor
$C^{[70,^28]}_{S_{11}(1535)}\simeq 1.10$ and
$C^{[70,^28]}_{S_{11}(1650)}\simeq -0.22$, which are compatible with
the previous quark model
determinations~\cite{Saghai:2001yd,He:2008ty,He:2009zzi}. These
determined values lead to a mixing angle $\theta_S\simeq
24^\circ\sim 32^\circ$. It is consistent with the mixing angles
determined by the partial decay widths of $S_{11}(1535)$ and
$S_{11}(1650)\to \pi N$ and $\eta N$ recently~\cite{liu-zhao-zhong}.
In Refs.~\cite{Saghai:2001yd,Saghai:2009zz}, the relation of
Eq.~(\ref{mix-coeff1}) was also obtained. However, the extracted
mixing angle $\theta_S = -35^\circ\sim -27^\circ$ appears to have a
sign difference from ours. This question has been clarified in their
recent work~\cite{An:2011sb}. In
Refs.~\cite{Saghai:2001yd,Saghai:2009zz,He:2008ty,He:2009zzi}, the
old conventions of the SU(3) wave functions from Isgur and Karl's
early works~\cite{Isgur:1977ef,Isgur:1978xj} were adopted, which
resulted in $R_S^N=1$. Thus, a negative value for the mixing angle
$\theta_S$ was obtained. We suggest one to adopt the Isgur's later
conventions of the SU(3) wave functions~\cite{Koniuk:1979vy} in the
studies, with which the relative signs among the resonances
transition amplitudes can be avoided. In line with the Isgur's later
conventions, $R_S^N=-1$ can be determined such that a positive value
for the mixing angle $\theta_S$ can be extracted.

Equation~(\ref{mix-coeff1}) provides a rather general relation for
the $[70,^28, 1/2^-]$ and $[70,^28, 1/2^-]$ mixings in $\gamma p\to
\eta p$. It has also an interesting implication of the mixing angle
range in terms of the experimental observations. Since so far the
experimental data suggests the dominance of $S_{11}(1535)$ near
threshold, the absolute value of
$|C^{[70,^28]}_{S_{11}(1535)}/C^{[70,^28]}_{S_{11}(1650)}|>>1$. It
should be noted that the positive mixing angle $\theta_S$ is
consistent with the OPE models~\cite{Chizma:2002qi}, $1/N_c$
expansion approach~\cite{Pirjol}, and the $^3P_0$ pair creation
model~\cite{Fizika}. The OGE model gives a negative value for the
mixing angle which is also natural since in the OGE model the
interacting vertex is defined differently~\cite{Isgur:1978xj}.

Similarly, $D_{13}(1520)$ and $D_{13}(1700)$ as mixing states of
$|70,^28, \frac{3}{2}^-\rangle$ and $|70,^48, \frac{3}{2}^-\rangle$
are expressed as follows:
\begin{eqnarray}
D_{13}(1520)= \cos\theta_D |70,^28, 3/2^-\rangle-\sin \theta_D |70,^48, 3/2^-\rangle ,\nonumber\\
D_{13}(1700)= \sin\theta_D |70,^28, 3/2^-\rangle + \cos \theta_D
|70,^48, 3/2^-\rangle,
\end{eqnarray}
where $\theta_D$ is the mixing angle. The configuration mixing
coefficient can be extracted as follows:
\begin{eqnarray}\label{mix-coeffD}
C^{[70,^28]}_{D_{13}(1520)}&=&R\cos\theta_D(\cos\theta_D-\sin\theta_D/\sqrt{10}),\nonumber\\
C^{[70,^28]}_{D_{13}(1700)}&=&R\sin\theta_D(\sin\theta_D+\cos\theta_D/\sqrt{10}),
\end{eqnarray}
where the factor $1/\sqrt{10}$ is given by $R_D^N\equiv \langle
N|H_m|70,^48, 3/2^-\rangle /\langle N|H_m|70,^28, 3/2^-\rangle
=1/\sqrt{10}$ in the $\eta$ production. Parameter $R$ is introduced
to adjust the overall $D$-wave strength. In $\gamma p\to \eta p$ due
to the absence of the $\langle p |H_e|70,^48 \rangle = 0$ helicity
amplitude at leading order, the amplitude of $\langle p |H_e|70,^28
\rangle$ cannot be well constrained. This amplitude will be canceled
out when taking the ratio of these two $D$-wave excitation
amplitudes. The deviation of $R$ from unity may imply the
underestimate of the EM transition amplitudes $\langle p |H_e|70,^28
\rangle$ in the symmetric quark model. In fact, it shows that the
$D_{13}(1520)$ helicity coupling $A^p_{3/2}\simeq 0.095$
GeV$^{-1/2}$ is underestimated by about a factor of $1.8$ compared
with the PDG value~\cite{Amsler:2008zzb}. In $\gamma p\to\eta p$,
the data favor $C_{D_{13}(1520)}\simeq 1.80$ and
$C_{D_{13}(1700)}\simeq 0.50$, which leads to a mixing angle
$\theta_D\simeq 21^\circ$ and $R\simeq 2.4$. The extracted mixing
angle is comparable with that extracted from the OPE
models~\cite{Chizma:2002qi}, $1/N_c$ expansion
approach~\cite{Pirjol}, $^3P_0$ pair creation model~\cite{Fizika},
and the chiral constituent quark approach~\cite{HeSaghai}.

In Figs.~\ref{fig-5-diff},~\ref{fig-5-diw} and \ref{fig-5-tot}, we
plot the differential cross sections, fixed-angle excitation
functions, and total cross section for $\gamma p\to \eta p$. These
observables can be well described with the parameters determined by
fitting 560 data points of the differential cross sections
from~\cite{McNicoll:2010qk} in the energy region 0.748 GeV $\leq
E_\gamma\leq$ 1.3 GeV. The $\chi^2$ per datum point is about
$\chi^2/N=2.9$. The dominant low-lying resonances are
$S_{11}(1535)$, $S_{11}(1650)$, $D_{13}(1520)$, $D_{13}(1700)$ and
$P_{13}(1720)$. In Tab.~\ref{csq} we can see that the value of
$\chi^2/N$ increases drastically if we switch off any one of those
resonances. This is to show a reasonable control of the model
parameters near threshold. The well determined amplitudes are then
applied to the quasi-free reaction with the deuteron target.

\begin{widetext}
\begin{center}
\begin{table}[ht]
\caption{The $\chi^2$ per datum point for the $\gamma p\rightarrow
\eta p$ (free) and $\gamma n\rightarrow \eta n$ (quasi-free). To
quantify the role played by different resonances, the values after
removing the corresponding resonances are also listed. } \label{csq}
\begin{tabular}{|c|c|c|c|c|c|c|c|c|c|c|c| }\hline\hline
              & full model & $S_{11}(1535)$ & $S_{11}(1650)$ &$D_{13}(1520)$ & $D_{13}(1700)$
              & $D_{15}(1675)$& $P_{13}(1720)$   \\\hline
$\chi^2/N$ ($\gamma p\rightarrow \eta p$)& 2.9   & 240     &  73.8  &7.3           & 4.4      & -- & 4.1      \\
$\chi^2/N$ ($\gamma n\rightarrow \eta n$)& 1.6   & 120  &5.1   &2.3   & 1.6   & 3.3  &4.0   \\
\hline
\end{tabular}
\end{table}
\end{center}
\end{widetext}

\subsection{$\gamma p\to \eta p$ with quasi-free proton target}

\begin{figure}[ht]
\centering \epsfxsize=8.5 cm \epsfbox{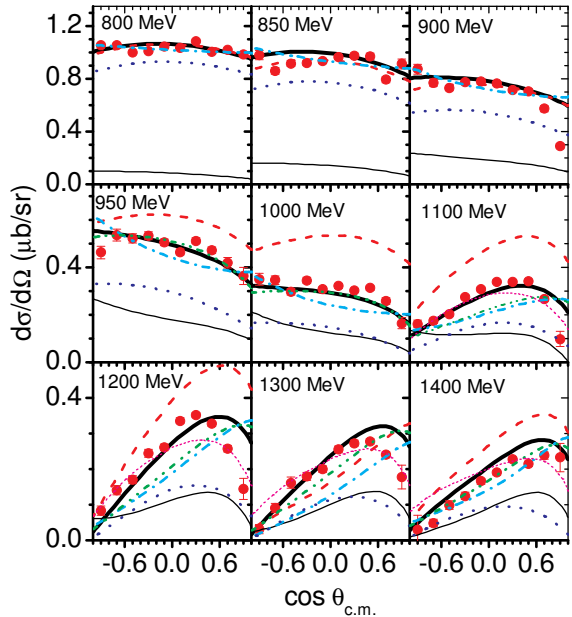} \caption{(Color
online) The differential cross sections for the $\eta$
photoproduction off the quasi-free proton at various beam energies.
The data are taken from~\cite{Jaegle:2008ux}. The solid curves
correspond to the full model result. The thin solid, dashed,
dash-dotted, dash-dot-dotted, short dashed, dotted curves are for
the results by switching off the contributions from $S_{11}(1535)$,
$S_{11}(1650)$, $D_{13}(1520)$, $D_{13}(1700)$, $P_{13}(1720)$, $u$
channel, respectively. }\label{fig-5}
\end{figure}

Taking into account the Fermi motion, our predictions for the
differential cross sections for the $\eta$ photo-production off the
quasi-free proton are shown in Fig.~\ref{fig-5}. It shows  that the
predictions are in a good agreement with the data from threshold to
$E_{\gamma}\simeq 1.4$ GeV. In this energy region, the
$S_{11}(1535)$ dominates the reaction, and $D_{13}(1520)$ is crucial
for accounting for the shape deviation from the $S$ wave. The $u$
channel also has effects on the differential cross section. By
switching off the amplitudes of $S_{11}(1535)$ (thin solid curves),
or $D_{13}(1520)$ (dash-dotted curves), or $u$ channel (dotted
curves), we see that the differential cross sections change
drastically.

Similar to the free nucleon reaction, the state mixings between
$[70, ^28]$ and $[70, ^48]$ are important. For $S_{11}(1535)$ and
$S_{11}(1650)$, their mixing angle is fixed in the free nucleon
reaction. In this sense, the quasi-free reaction can be treated as a
test of the mixing scenario near threshold. The relative destructive
interference between $S_{11}(1535)$ and $S_{11}(1650)$ accounts for
the sudden change of the angular distributions around
$E_\gamma=1.00$ GeV. In this energy region, the dashed curves
indicate that without the contributions of $S_{11}(1650)$ the
differential cross sections would be enhanced significantly. It
should be mentioned that the destructive interference between
$S_{11}(1535)$ and $S_{11}(1650)$ also accounts for the dip
structure around $W = 1670$ MeV in the $\gamma p\rightarrow \eta p$
excitation function at forward angles observed by MAMI-C
Collaboration recently (see Fig.~\ref{fig-5-diw}).

It can be also recognized that $D_{13}(1700)$ plays an important
role around $E_\gamma=(1.0\sim 1.1)$ GeV. We find that a sizable
strength of $D_{13}(1700)$, i.e. $ C_{D_{13}(1700)}=0.50$, is needed
in order to account for the angular distributions. According to our
analysis, $D_{13}(1700)$ is the main contributor to the bump
structure around $W=1.7$ GeV recently observed by MAMI-C
Collaboration in the cross section of the $\eta$ production off free
proton (See Fig.~\ref{fig-5-tot}). It should be pointed out that in
the present work we include sizeable contributions from
$D_{13}(1700)$ to improve the agreement with the data for the
differential cross sections, which is compatible with the suggestion
of Nakayama \emph{et al.}~\cite{Nakayama:2008tg}. Another possible
explanation was discussed within the quark
model~\cite{Li:1998ni,Saghai:2001yd,He:2008ty,He:2009zzi}, where a
new $S_{11}$ with a mass of $\sim 1.72$ GeV was suggested to be
included to improve the descriptions of the differential cross
sections.

The $P$-wave state $P_{13}(1720)$ turns out to have rather small
effects on the cross section near threshold, but shows up in the
region of $E_\gamma =1.2\sim 1.4$ GeV in the forward angle direction
shown by the short dashed lines. Some contributions of
$P_{13}(1720)$ in the $\gamma p\to \eta p$ reaction were also
suggested in~\cite{Shyam:2008fr,He:2008ty,Nakayama:2008tg}.

The $u$ channel  plays an important role in the reactions. The
dotted curves show that by turning off the contributions of $u$
channel backgrounds, the cross sections will be underestimated
significantly.

\begin{center}
\begin{figure}[ht]
\centering \epsfxsize=8.5 cm \epsfbox{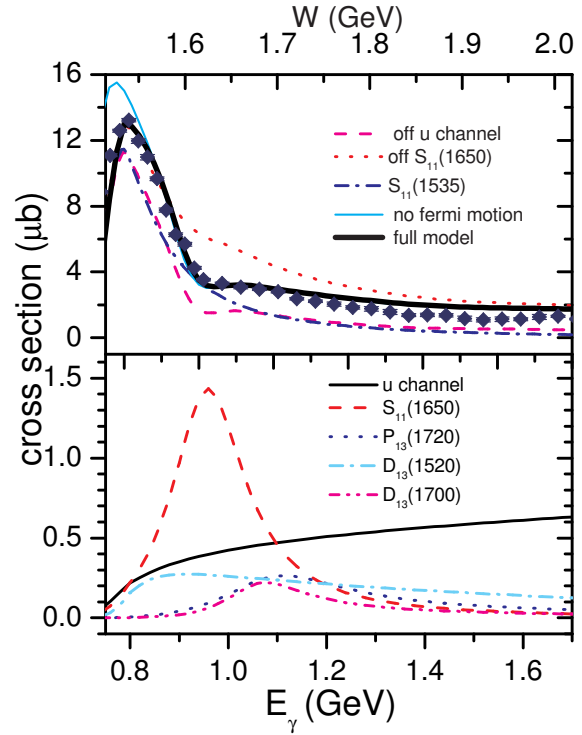} \caption{ (Color
online) The quasi-free cross section for $\gamma p\rightarrow \eta
p$. The data are taken from~\cite{Jaegle:2008ux}. In the upper panel
the bold solid curve corresponds to the full model result, while the
dashed, dotted curves are for the results by switching off the
contributions from the $u$ channel and $S_{11}(1650)$, respectively.
The partial cross sections for $S_{11}(1535)$, $S_{11}(1650)$,
$D_{13}(1520)$, $D_{13}(1700)$, $P_{13}(1720)$, $u$ channel are
indicated explicitly by different legends in the
figure.}\label{fig-6}
\end{figure}
\end{center}

In Fig.~\ref{fig-6}, the total cross section and the exclusive cross
sections of several single resonances are shown. Our results are in
good agreement with the data up to $E_\gamma\simeq 1.8$ GeV. The
thin line in the upper panel of Fig.~\ref{fig-6} denotes the result
of no Fermi motion, i.e. the cross section for the free proton
reaction in Fig.~\ref{fig-5-tot}. It shows that the Fermi motion has
the most significant corrections around $S_{11}(1535)$ energy region
and becomes negligible in the higher energy region.

Apart from the Fermi motion corrections, the main feature of the
$\eta$ photoproduction off quasi-free protons is similar to the case
of free proton reaction. In particular, we still see the dominance
of $S_{11}(1535)$ near threshold and its destructive interference
with $S_{11}(1650)$ which accounts for a shallow dip near
$E_\gamma\simeq 1.05$ GeV ($W\simeq 1.68$ GeV).

\begin{center}
\begin{figure}[ht]
\centering \epsfxsize=9.0 cm \epsfbox{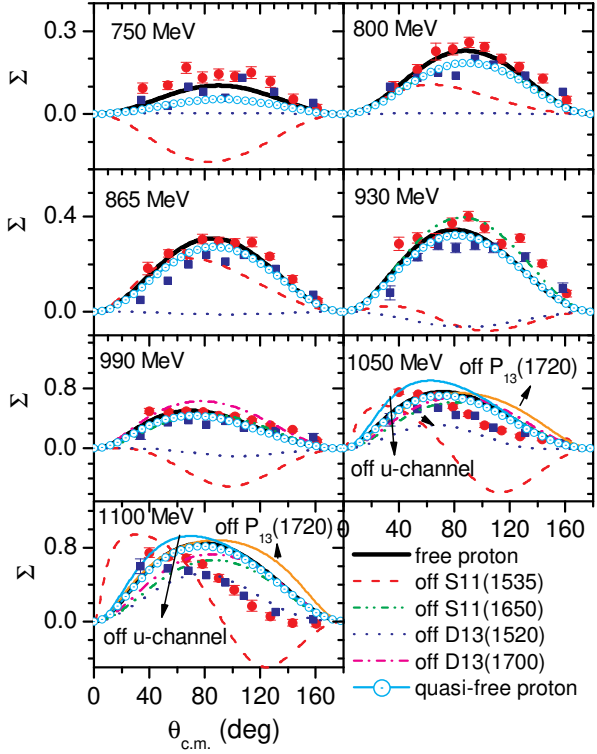} \caption{ (Color
online) Beam asymmetry $\Sigma$ in the $\eta$ photoproduction off
quasi-free protons as a function of $\theta_{c.m.}$. The solid
squares correspond to the quasi-free proton data from
Ref.~\cite{Fantini:2008zz}, while the solid circles stand for the
free proton data from Ref.~\cite{Bartalini:2007fg}. The bold solid
curves are the full model results. The results by switching off the
contributions from $S_{11}(1535)$, $S_{11}(1650)$, $D_{13}(1520)$,
$D_{13}(1700)$, $P_{13}(1720)$, and $u$ channel are indicated
explicitly by different legends in the figure. }\label{fig-7}
\end{figure}
\end{center}

Polarization observables should be more sensitive to the underlying
mechanisms. In Fig.~\ref{fig-7}, we plot the polarized beam
asymmetry for $\gamma p\rightarrow \eta p$ in comparison with the
data for both quasi-free (solid circles) and free (open circles)
proton reactions. These two sets of data exhibit similarities with
each other, which indicate the small effects from the Fermi motion
corrections. The theoretical results are able to reproduce the data
agreeably up to $E_\gamma\simeq 1.1$ GeV. The discrepancies come up
in the higher energy region of $E_\gamma> 1.05$ GeV, where the
contributions from higher resonances may become important.

By switching off the single resonance contributions, we can examine
their roles in the reaction. It shows that the large asymmetries in
the intermediate angle is due to the dominant $S$-wave interferences
with the $D$-wave.

The interferences of $S_{11}(1535)$ with $D_{13}(1520)$ dominate the
beam asymmetry. As shown by the dashed and dotted curves, if we
switch off the contributions of $S_{11}(1535)$ and $D_{13}(1520)$,
respectively, the asymmetries change drastically. The important
roles of $S_{11}(1535)$ and $D_{13}(1520)$ in the beam asymmetry are
also suggested
in~\cite{He:2009zzi,He:2008ty,Chiang:2002vq,Tiator:1999gr}.

The contributions of $S_{11}(1650)$, $D_{13}(1700)$, $P_{13}(1720)$
and $u$ channel backgrounds begin to appear in the energy region
$E_\gamma> 1.0$ GeV. With the increasing energy the contributions
become more and more obvious. $P_{13}(1720)$ seems to be responsible
for the shifting of the maximum to forward angles. If we switch off
its contributions, the beam asymmetry will keep an approximate
forward-backward symmetry.

In brief, it shows that the configuration mixing effects are
important for describing $\gamma p\rightarrow \eta p$ near
threshold. $S_{11}(1535)$ dominates the cross section, and its
destructive interference with $S_{11}(1650)$ naturally accounts for
the dip structure around $W\simeq 1.68$ GeV in the total cross
section. The $D$-wave state, i.e. $D_{13}(1520)$, is crucial for
accounting for the angular distributions in the differential cross
sections and beam asymmetry as found by other
analyses~\cite{Shyam:2008fr,Chiang:2002vq,Chiang:2001as,Tiator:1999gr,Knochlein:1995qz}.
Signals for $D_{13}(1700)$, $P_{13}(1720)$ and $u$-channel
contributions are also evidential.

\subsection{$\gamma n\rightarrow \eta n$ with quasi-free neutron target}

\begin{center}
\begin{figure}[ht]
\centering \epsfxsize=9.0 cm \epsfbox{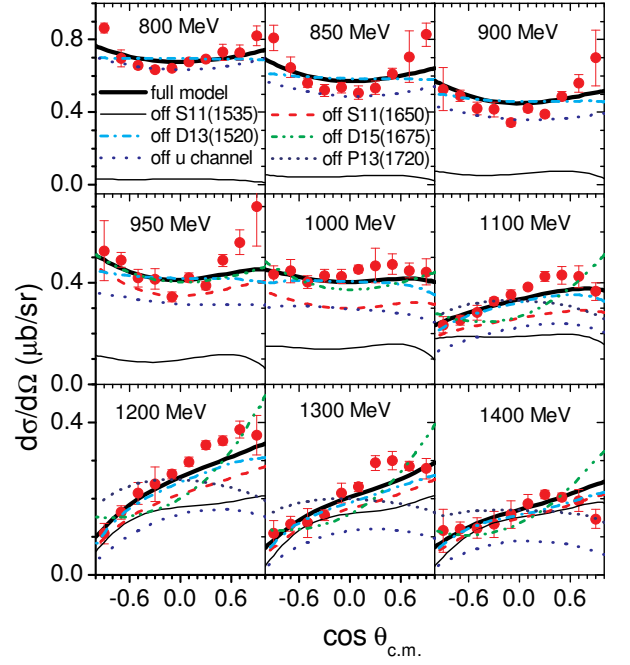} \caption{(Color
online) The differential cross sections for the $\eta$
photoproduction off the quasi-free neutron at various beam energies.
The data are taken from Ref.~\cite{Jaegle:2008ux}. The solid curves
correspond to the full model results. The thin solid, dashed,
dash-dotted, dash-dot-dotted, short dashed, dotted curves are for
the results by switching off the contributions from $S_{11}(1535)$,
$S_{11}(1650)$, $D_{13}(1520)$, $D_{15}(1675)$, $P_{13}(1720)$, $u$
channel, respectively. }\label{cmf}
\end{figure}
\end{center}

\begin{center}
\begin{figure}[ht]
\centering \epsfxsize=7.0 cm \epsfbox{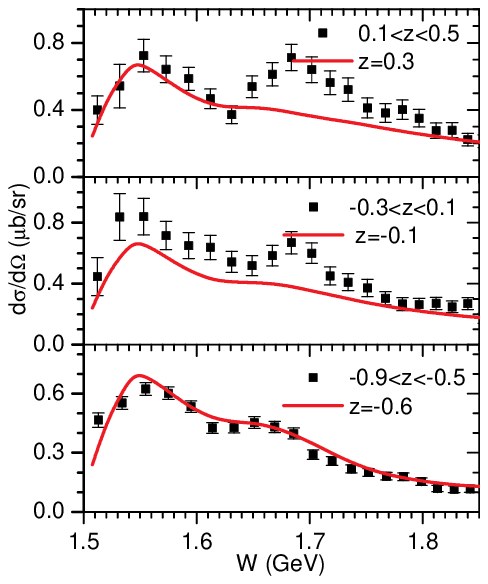} \caption{ (Color
online) Fixed-angle excitation functions as a function of center
mass energy $W$ for the quasi-free $\gamma n\rightarrow \eta n$
process. Where $z$ is defined by $z=\cos \theta_{c.m.}$. The data
are taken from \cite{Kuznetsov:2007gr}.}\label{fig-2}
\end{figure}
\end{center}

For the $\gamma n\rightarrow \eta n$ process, the Moorhouse
selection rule no longer applies. Thus, both the spin-3/2 and
spin-1/2 resonances would contribute in the $s$ channel. Apart from
the inevitable Fermi motion corrections, another feature with
$\gamma n\rightarrow \eta n$ is that the EM interaction only
involves neutral hadrons. Therefore, the electric terms would vanish
and the leading EM coupling would come from the magnetic terms. As a
consequence, the neutron reaction does not share the same resonance
strength parameters $C_R$ with the proton reaction, although the
strong interaction vertices can be well connected with each other by
the isospin symmetry relation. With the same mixing scheme in
Eq.(\ref{mix-s11}) we can extract the configuration mixing
coefficients for $S_{11}(1535)$ and $S_{11}(1650)$, which are
\begin{eqnarray}\label{mix-coeff}
C^{[70,^28]}_{S_{11}(1535)}&=&R^S_2\cos\theta_S(\cos\theta_S+\sin\theta_S), \\
C^{[70,^48]}_{S_{11}(1535)}&=&R^S_4\sin\theta_S(\sin\theta_S+\cos\theta_S),\\
C^{[70,^28]}_{S_{11}(1650)}&=&R^S_2\sin\theta_S(\sin\theta_S-\cos\theta_S), \\
C^{[70,^48]}_{S_{11}(1650)}&=&R^S_4\cos\theta_S(\cos\theta_S-\sin\theta_S),
\end{eqnarray}
where $R^S_2$ and $R^S_4$ are introduced to adjust the strength of
the partial wave amplitudes of $[70,^28,1/2^-]$ and
$[70,^48,1/2^-]$, respectively, which may be overestimated or
underestimated in the naive quark model.

Furthermore, considering the configuration mixing effects in the
$D_{13}(1520)$ and $D_{13}(1700)$, we extract the configuration
mixing coefficient as follows:
\begin{eqnarray}\label{mix-coeffD1}
C^{[70,^28]}_{D_{13}(1520)}&=&R^D_2\cos\theta_D(\cos\theta_D-\frac{1}{\sqrt{10}}\sin\theta_D), \\
C^{[70,^48]}_{D_{13}(1520)}&=&R^D_4\sin\theta_D(\sin\theta_D-\sqrt{10}\cos\theta_D),\\
C^{[70,^28]}_{D_{13}(1700)}&=&R^D_2\sin\theta_D(\sin\theta_D+\frac{1}{\sqrt{10}}\cos\theta_D), \\
C^{[70,^48]}_{D_{13}(1700)}&=&R^D_4\cos\theta_D(\cos\theta_D+\sqrt{10}\sin\theta_D),
\end{eqnarray}
where $R^D_2$ and $R^D_4$ are the constants which are introduced to
adjust the strength of the partial wave amplitudes of
$[70,^28,3/2^-]$ and $[70,^48,3/2^-]$, respectively.

\begin{center}
\begin{figure}[ht]
\centering \epsfxsize=8.0 cm \epsfbox{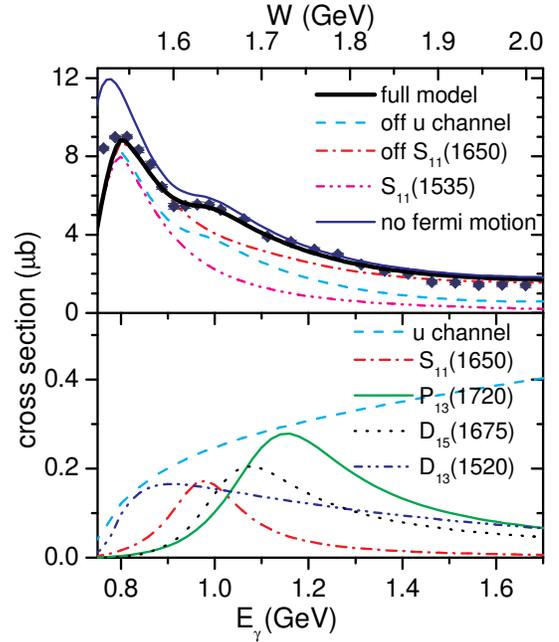} \caption{ (Color
online) The cross section for the quasi-free $\gamma n\rightarrow
\eta n$ reaction. The data are from Ref.~\cite{Jaegle:2008ux}. In
the upper panel the heavy (thin) solid curve corresponds to the full
model result with (without) Fermi motion corrections, while the
dashed, dash-dotted curves are for the results by switching off the
contributions from the $u$ channel and $S_{11}(1650)$, respectively.
The partial cross sections for $S_{11}(1535)$, $S_{11}(1650)$,
$D_{13}(1520)$, $D_{15}(1675)$, $P_{13}(1720)$, $u$ channel are
indicated explicitly by different legends in the
figure.}\label{fig-f3}
\end{figure}
\end{center}

\begin{center}
\begin{figure}[ht]
\centering \epsfxsize=8.0 cm \epsfbox{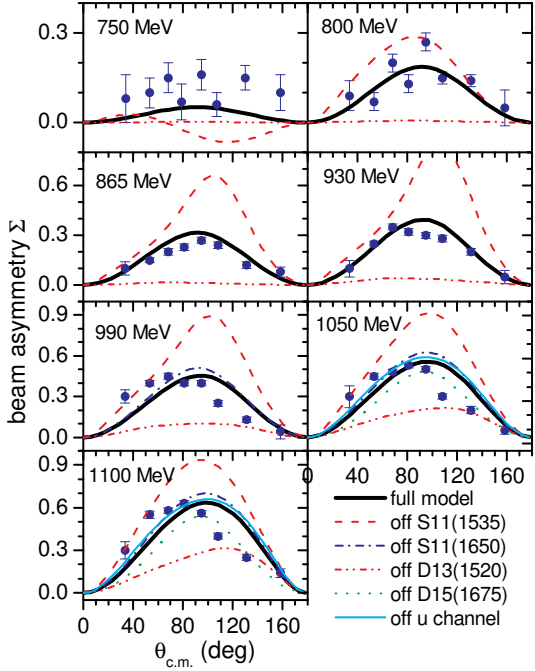} \caption{ (Color
online) Beam asymmetry $\Sigma$ in $\eta$ photoproduction off the
quasi-free neutron as a function of $\theta_{c.m.}$. The data are
taken from Ref.~\cite{Fantini:2008zz}. The solid curves are the full
model results. The results by switching off the contributions from
$S_{11}(1535)$, $S_{11}(1650)$, $D_{13}(1520)$, $D_{15}(1675)$, and
$u$ channel are indicated explicitly by different legends in the
figure.}\label{fig-f4}
\end{figure}
\end{center}

In this reaction the mixing angles,  $\theta_S=25^\circ$ and
$\theta_D=20^\circ$ are fixed  in the $\gamma p\rightarrow \eta p$
process. The $R$ values ($R^S_2$, $R^S_4$, $R^D_2$ and $R^D_4$) and
$C_R$ values for the $D_{15}$ and $P_{13}$ resonances are determined
by fitting the 90 data points of differential cross sections in the
energy region 0.8 GeV$\leq E_\gamma\leq$1.4 GeV with $\chi^2/N=1.6$.
With these fitted values, $R^S_2=0.71$, $R^S_4=2.95$, $R^D_2=1.55$
and $R^D_4=1.0$, we obtain a good description of the differential
cross sections, beam asymmetries, and total cross section from
threshold up to $E_{\gamma}\simeq 1.2$ GeV as shown in
Figs.~\ref{cmf}-\ref{fig-f4}. With these determined $R$ values, the
$C_R$ values for the $S$ and $D$ wave resonances are extracted and
listed in Tab.~\ref{param C}.

In $\gamma p \to\eta p$ the main resonance contributions are from
$S_{11}(1535)$, $S_{11}(1650)$, $D_{13}(1520)$, $D_{15}(1675)$ and
$P_{13}(1720)$. As shown in Tab.~\ref{csq}, the $\chi^2/N$ value
would increase drastically if any one of these resonances is absent
in the fitting. Among the main contributors, the $S_{11}(1535)$
plays a dominant role as well-known in the literature. If we turn
off its contribution (see the thin solid line in Fig.~\ref{cmf}),
the differential cross sections become tiny, and the $\chi^2/N$
increases to a very large value of 120. The constructive
interferences between $S_{11}(1650)$ and $S_{11}(1535)$ around
$E_{\gamma}\simeq (1.1\pm 0.2)$ GeV can be clearly seen in both
differential and total cross sections. Turning off the contributions
of $S_{11}(1650)$ (see the dashed curves the Fig.~\ref{cmf}) the
cross section is apparently underestimated. With the constructive
interferences between $S_{11}(1535)$ and $S_{11}(1650)$, we can
naturally explain the second small bump-like structure at $W\simeq
1.68$ GeV ($E_{\gamma}\simeq 1.0$ GeV) in the total cross section
and excitation function (see Figs.~\ref{fig-2} and \ref{fig-f3}),
although the theoretical results has slightly underestimated the
data for the excitation functions at forward angles. Our conclusion
is compatible with the recent partial-wave analysis
results~\cite{Anisovich:2008wd} and the coupled-channel K-matrix
approach~\cite{Shyam:2008fr}. The dash-dotted line in the upper
panel of Fig.~\ref{fig-2} shows that the second bump disappears as
long as $S_{11}(1650)$ is switches off. Furthermore, we noted that
there are other explanations about the bump-like structure at
$W\simeq 1.68$ GeV. For example, Shklyar \emph{et al.} explained the
bump-like structure in terms of coupled-channel effects due to
$S_{11}(1650)$ and $P_{11}(1710)$~\cite{Shklyar:2006xw}. In
Refs.~\cite{Choi:2005ki, Fix:2007st}, the structure around $W=1.68$
GeV was even considered as evidence for a non-strange member of the
anti-decuplet pentaquark state $P_{11}(1670)$~\cite{Jaffe:2003sg}.

$D_{13}(1520)$ is crucial for the shape change of the differential
cross sections in the low energy region $E_{\gamma}< 1.0$ GeV,
although its effects on the total cross section are negligible. The
dash-dotted curves in Fig.~\ref{cmf} show that without its
contributions, the forward angle enhancement would disappear. The
importance of $D_{13}(1520)$ can also be seen in the beam asymmetry.
In Fig.~\ref{fig-f4} it shows that the interferences between
$D_{13}(1520)$ and $S_{11}(1535)$ govern the beam asymmetry at the
low energy region $E_{\gamma}< 1.1$ GeV. Switching off the
$D_{13}(1520)$ (dash-dot-dotted curves ) or $S_{11}(1535)$ (dashed
curves), we find that the curves change drastically.

A relatively large contribution from $D_{15}(1675)$ with
$C_{D_{15}(1675)}=1.8$ seems to be favored by the experimental data.
In Fig. \ref{cmf}, as indicated by the dash-dot-dotted lines, the
cross sections at forward angles are overestimated significantly
without $D_{15}(1675)$. The relatively large strength parameter,
$C_{D_{15}(1675)}=1.8$, for $D_{15}(1675)$ might due to the
underestimate of the helicity couplings in the naive quark model. In
the SU(6)$\otimes$O(3) symmetry limit, the helicity couplings of
$D_{15}(1675)$ are $A^n_{1/2}\simeq -0.032$ GeV$^{-1/2}$ and
$A^n_{3/2}\simeq -0.045$ GeV$^{-1/2}$, which are much smaller than
the center values from the PDG~\cite{Amsler:2008zzb}.

In the SU(6)$\otimes$O(3) symmetry limit, the contributions of
$P_{13}(1720)$ are too small to reproduce the forward peak in the
differential cross sections in the energy region $E_{\gamma}\simeq
(1.2\pm 0.1)$ GeV. By fitting the data, it shows that a relatively
large value $C_{P_{13}(1720)}=3.0$ should be applied. As denoted by
the short-dashed line in Fig. \ref{cmf} we see that without
$P_{13}(1720)$, the forward peak in the differential cross sections
would disappear. Contributions from $P_{13}(1720)$ were also
suggested by Klempt \emph{et al.}~\cite{Anisovich:2008wd}. The
unexpectedly large strength parameter of $P_{13}(1720)$ in the
reaction could be an indication that the SU(6)$\otimes$O(3) symmetry
is badly violated~\cite{Zhao:2006an}.

The $u$ channel background also plays an important role in the
$\gamma n\rightarrow \eta n$ process. It has a constructive
interferences with $S_{11}(1535)$. As shown by the dotted lines in
Fig.~\ref{cmf} and dashed line in the upper panel of
Fig.~\ref{fig-f3}, the cross sections will be underestimated more
and more obviously with the increasing energy if the $u$ channel is
switched off.

In the lower panel of the Fig. \ref{fig-f3}, it shows that the
$D_{13}(1520)$, $D_{15}(1675)$ and $P_{13}(1720)$ and background $u$
channel have comparable exclusive cross sections with $S_{11}(1650)$
around $E_{\gamma}\simeq 1.1$ GeV. Their overall interfering effects
have been essential for a good description of the experimental
observable in this energy region. For instance, it can be noticed
that the angular distributions of the cross sections are sensitive
to $D_{13}(1520)$, $D_{15}(1675)$ and $P_{13}(1720)$, although their
effects on the total cross section are negligible. Furthermore, the
$S_{11}(1650)$, $D_{15}(1675)$ and background $u$ channel have also
sizeable effects on the beam asymmetry. Comparing with the data, we
find that the predicted maximum for the beam asymmetry slightly
shifts to the backward angles in the region $E_{\gamma}\gtrsim 1.0$
GeV. It might indicate that higher resonances have interferences and
they should be included with more elaborate considerations.

We also mention that other resonances, such as $P_{11}(1440)$,
$P_{11}(1710)$, $D_{13}(1700)$, have negligible effects in the
$\eta$ photoproduction.

Similar to the $\gamma p\rightarrow \eta p$ process with quasi-free
protons, the Fermi motion brings corrections to the cross section
from threshold up to the mass of $S_{11}(1535)$. The effects
decrease with the increasing energy as denoted by the thin solid
curve in the upper panel of Fig.~\ref{fig-f3}.

To briefly summarize this subsection, we show that with the
well-established $S_{11}(1535)$, $S_{11}(1650)$, $D_{13}(1520)$,
$D_{15}(1675)$ and $P_{13}(1720)$ resonances and contribution of the
$u$ channel, we can obtained a good description of the reaction
$\gamma n\rightarrow \eta n$ with quasi-free neutrons. In this
reaction, $S_{11}(1535)$ plays a dominated role similar to $\gamma
p\to \eta p$. However, due to the presence of $[70, ^48]$ states,
the amplitude of $S_{11}(1535)$ appears to have a constructive
interference with $S_{11}(1650)$, which is crucial to explain the
bump-like structure around $W=1.68$ GeV. We also find that the
angular distributions of the differential cross sections are
sensitive to $D_{15}(1675)$ and $P_{13}(1720)$ around $E_\gamma=1.1$
GeV, and the $u$ channel contributes a large background to the
differential cross sections. The interferences between
$D_{13}(1520)$ and $S_{11}(1535)$ govern the beam asymmetry at the
low energy region $E_{\gamma}< 1.1$ GeV, while contributions of
several other resonances, such as $P_{11}(1440)$, $P_{11}(1710)$ and
$D_{13}(1700)$, are negligibly small.

\subsection{$\sigma_n/\sigma_p$ with quasi-free neutron target}

\begin{center}
\begin{figure}[ht]
\centering \epsfxsize=8.0 cm \epsfbox{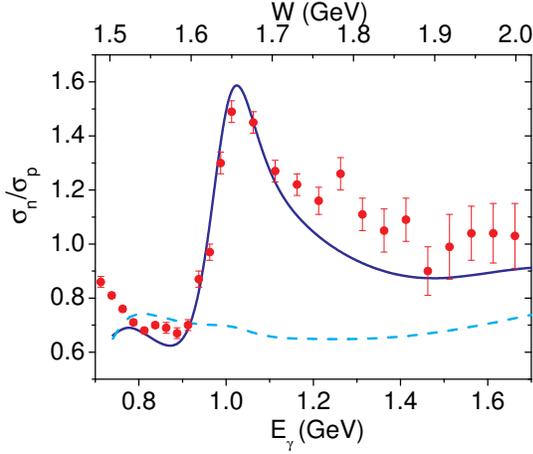} \caption{ (Color
online) Cross section ratio $\sigma_n/\sigma_p$ for the quasi-free
reactions $\gamma n\rightarrow \eta n$ and $\gamma p\rightarrow \eta
p$ as a function of the photon energy. The data are taken from
Ref.~\cite{Jaegle:2008ux}. The dashed curve is the result with the
$S_{11}(1650)$ switched off. }\label{fig-8}
\end{figure}
\end{center}

With the reasonable description of the $\eta$ photoproduction off
the free nucleons available, we can proceed a prediction for the
quasi-free neutron/proton cross section ratio, $\sigma_n/\sigma_p$.
This quantity was measured by CBELSA/TAPS
Collaboration~\cite{Jaegle:2008ux}. In Fig.~\ref{fig-8} the
theoretical prediction is plotted as the solid curve to compare with
the experimental measurement~\cite{Jaegle:2008ux}. It shows that the
data can be well described by the solid curve, and an enhancement
around $W\simeq 1.68$ GeV is observable.

In the literature this sharp peak at $W\simeq 1.68$ GeV is
considered as an evidence of a new resonance
~\cite{Kuznetsov:2004gy,Miyahara:2007zz,Werthmuller:2010af,
Kuznetsov:2010as,Fix:2007st,Choi:2005ki,Kuznetsov:2007gr}. However,
in our case it can be naturally interpreted as the interferences
between $S_{11}(1650)$ and $S_{11}(1535)$, i.e. destructive in
$\gamma p\to \eta p$ but constructive in $\gamma n\to \eta n$. The
dashed curve in Fig.~\ref{fig-8} denotes the cross section ratio
with $S_{11}(1650)$ switched off. It shows that the ratio becomes
flat and the sharp peak disappears. Our prediction agrees with that
of the recent partial-wave analysis~\cite{Anisovich:2008wd}. The
ratio is also studied within a unitary coupled channel model, it is
interpreted by the coupled channel effect in $S$ waves, where the
$S_{11}(1535)$ is dynamically generated~\cite{Doring:2009qr}.

\subsection{helicity amplitudes }

\begin{table}[ht]
\caption{The expressions of $\xi$ in Eq.~\ref{hlc} for various
resonances. Where we have defined
$\mathcal{\mathcal{K}}\equiv\sqrt{\alpha_e\alpha_\eta
\pi(M_N+E_f)/M_R^3}/\Gamma_R$, $\mathcal{A}\equiv
\left[\frac{2\omega_\gamma}{m_q}-\frac{2q^2}{3\alpha^2}(1+\frac{\omega_\eta}{E_f+M_N})
\right]e^{-\frac{k^2+q^2}{6\alpha^2}}$, and
$\mathcal{B}=-\frac{2q^2}{3\alpha^2}(1+\frac{\omega_\eta}{E_f+M_N})e^{-\frac{k^2+q^2}{6\alpha^2}}$.
}\label{hlx}
\begin{tabular}{|c c|c| }\hline
 \hline $S_{11}(1535)$ & $\xi^p_{1/2}$ &
$\mathcal{K}\frac{\omega_\gamma}{6}(1+\frac{\omega_\gamma}{2m_q})
\mathcal{A}C^{[70,^28]p}_{S_{11}(1535)}$     \\
                       & $\xi^n_{1/2}$ & $-\mathcal{K}
\left[\frac{\omega_\gamma}{6}(1+\frac{\omega_\gamma}{6m_q})
C^{[70,^28]n}_{S_{11}(1535)}+\frac{\omega_\gamma^2}{36m_q}C^{[70,^48]n}_{S_{11}(1535)}
\right]\mathcal{A}$ \\ \hline $S_{11}(1650)$  & $\xi^p_{1/2}$ &
$-\mathcal{K}\frac{\omega_\gamma}{6}(1+\frac{\omega_\gamma}{2m_q})\mathcal{A}C^{[70,^28]p}_{S_{11}(1650)}$      \\
                &  $\xi^n_{1/2}$ & $\mathcal{K}
\left[\frac{\omega_\gamma}{6}(1+\frac{\omega_\gamma}{6m_q})
C^{[70,^28]n}_{S_{11}(1650)}+\frac{\omega_\gamma^2}{36m_q}C^{[70,^48]n}_{S_{11}(1650)}
\right]\mathcal{A}$     \\ \hline
        $D_{13}(1520)$     & $\xi^p_{1/2}$ & $-\mathcal{K}
\sqrt{\frac{1}{2}}\frac{\omega_\gamma}{6}(1-\frac{\omega_\gamma}{m_q})
\mathcal{B}C^{[70,^28]p}_{D_{13}(1520)}
$    \\
               &  $\xi^p_{3/2}$ & $\mathcal{K} \sqrt{\frac{1}{2}}\frac{\omega_\gamma}{6}
\mathcal{B}C^{[70,^28]p}_{D_{13}(1520)}$    \\
              &  $\xi^n_{1/2}$ & $-\sqrt{\frac{1}{2}}\mathcal{K}
\left[\frac{\omega_\gamma}{6}(1-\frac{\omega_\gamma}{3m_q})
C^{[70,^28]n}_{D_{13}(1520)}+\frac{\omega_\gamma^2}{180m_q}C^{[70,^48]n}_{D_{13}(1520)}
\right]\mathcal{B}$    \\
             &  $\xi^n_{3/2}$ & $-\sqrt{\frac{1}{2}}\mathcal{K}
\left[\frac{\omega_\gamma}{2\sqrt{3}}
C^{[70,^28]n}_{D_{13}(1520)}+\frac{\omega_\gamma^2}{20\sqrt{3}m_q}C^{[70,^48]n}_{D_{13}(1520)}
\right]\mathcal{B}$    \\ \hline
        $D_{13}(1700)$  & $\xi^n_{1/2}$&
$-\sqrt{\frac{1}{2}}\mathcal{K}
\left[\frac{\omega_\gamma}{6}(1-\frac{\omega_\gamma}{3m_q})
C^{[70,^28]n}_{D_{13}(1700)}+\frac{\omega_\gamma^2}{180m_q}C^{[70,^48]n}_{D_{13}(1700)}
\right]\mathcal{B}$\\
             &  $\xi^n_{3/2}$ & $-\sqrt{\frac{1}{2}}\mathcal{K}
\left[\frac{\omega_\gamma}{2\sqrt{3}}
C^{[70,^28]n}_{D_{13}(1700)}+\frac{\omega_\gamma^2}{20\sqrt{3}m_q}C^{[70,^48]n}_{D_{13}(1700)}
\right]\mathcal{B}$    \\
             &  $\xi^p_{1/2}$ & $-\mathcal{K}
\sqrt{\frac{1}{2}}\frac{\omega_\gamma}{6}(1-\frac{\omega_\gamma}{m_q})
\mathcal{B}C^{[70,^28]p}_{D_{13}(1700)}$\\
             &  $\xi^p_{3/2}$ & $\mathcal{K} \sqrt{\frac{1}{2}}\frac{\omega_\gamma}{6}
\mathcal{B}C^{[70,^28]p}_{D_{13}(1700)}$
\\ \hline
$D_{15}(1675)$&  $\xi^n_{1/2}$ & $-\mathcal{K}
\frac{\omega_\gamma^2}{40m_q} \mathcal{B}C_{D_{15}(1675)}
$    \\
            &  $\xi^n_{3/2}$ & $-\mathcal{K} \frac{\omega_\gamma^2}{20\sqrt{2}m_q}
\mathcal{B}C_{D_{15}(1675)}
$    \\
\hline
\end{tabular}
\end{table}

The helicity amplitudes of the nucleon resonances can be extracted
by the relation~\cite{Benmerrouche:1994uc},
\begin{eqnarray}\label{hlc}
A^{n,p}_{1/2,3/2}=\sqrt{\frac{|\mathbf{q}|M_R\Gamma_R}{|\mathbf{k}|M_Nb_\eta}}\xi^{n,p}_{1/2,3/2},
\end{eqnarray}
where $b_\eta\equiv \Gamma_{\eta N}/\Gamma_R$ is the branching ratio
of the resonance, which can be obtained from the experimental
determinations. The quantity $\xi$ for different resonances can be
analytically expressed from their CGLN amplitudes. In this work, we
have given the expressions of the $\xi$ for the low-lying $S$ and
$D$-wave resonances in Tab.~\ref{hlx}.

In the calculations we take the branching ratios of various
resonances
from~\cite{Amsler:2008zzb,Tiator:1999gr,Chiang:2001as,Batinic:1995kr},
which have been listed in Tab.~\ref{HL}. Together with the model
parameter fixed, we can extract the helicity amplitudes of those
low-lying $S$ and $D$-wave resonances that are listed in
Tab.~\ref{HL}. The PDG values~\cite{Amsler:2008zzb} are also listed
as a comparison. It is found that the predicted magnitudes for
$A_{1/2}$ and $A_{3/2}$ are compatible with the PDG values at the
30\% level.

It should be mentioned that when we take the small branching ratio
$b_\eta=0.05\pm 0.02\%$ from~\cite{Tiator:1999gr,Chiang:2001as} for
$D_{13}(1520)$, its extracted helicity amplitudes are good agreement
with the values from PDG. However, if we use the large branching
ratio $b_\eta=2.3\pm 0.4\%$ from PDG, the helicity amplitudes are
too small to be comparable with the experimental values.

Furthermore, we have noted that the $A^n_{1/2}$ for $S_{11}(1650)$
extracted by us has a positive sign, which is opposite to that of
other predictions~\cite{Anisovich:2008wd,Koniuk:1979vy,
Li:1990qu,Bijker:1994yr,Capstick:1992uc}. Without configuration
mixing between $[70, ^28]$ and $[70, ^48]$, the value of $A^n_{1/2}$
for $S_{11}(1650)$ predicted from quark model is positive, which
totally comes from $[70, ^48]$~\cite{Koniuk:1979vy,
Li:1990qu,Bijker:1994yr,Capstick:1992uc}. While, when the
configuration mixing effects are included, the previous quark model
study predicted a negative value of $A^n_{1/2}$ for $S_{11}(1650)$,
which is dominated by $[70, ^28]$~\cite{Koniuk:1979vy,
Li:1990qu,Bijker:1994yr,Capstick:1992uc}. In this work, we make some
corrections for the amplitudes of $[70, ^28]$ and $[70, ^48]$ to
reproduce the data for $\gamma n\rightarrow \eta n$ reaction. It is
found that a strong contribution of $[70, ^48]$ might be needed in
the reaction, which leads to a dominant contribution of $[70, ^48]$
to the $A^n_{1/2}$ in the configuration mixing scheme. Thus, we
obtain a positive sign.

The branching ratio $b_\eta$ for $D_{15}(1675)$ has a large
uncertainty. With the upper limit of $b_\eta$ (i.e. $b_\eta=0.01$),
the extracted helicity amplitudes, $A^n_{1/2}\simeq-40$ and
$A^n_{1/2}\simeq-56$, are good agreement with the values of PDG,
which indicates the branching ratio $b_\eta$ might favor a large
value of $b_\eta\simeq 0.01$.

\begin{widetext}
\begin{center}
\begin{table}[ht]
\caption{Estimated helicity amplitudes of the $S$ and $D$-wave
resonances (in $ 10^{-3}$GeV$^{-1/2}$).} \label{HL}
\begin{tabular}{|c|c|c||c|c||c|c||c|c|c|c|c|c|c| }\hline\hline
resonance& $A^p_{1/2}$  & $A^{p}_{1/2}$(PDG) & $A^p_{3/2}$
&$A^{p}_{3/2}$(PDG) & $A^n_{1/2}$  & $A^n_{1/2}$(PDG)& $A^n_{3/2}$ &
$A^n_{3/2}$ (PDG)& $b_\eta$ used in this work
\\\hline $S_{11}(1535)$& $60\pm 5$  & $90\pm30$ & -- &-- & $-68\pm
5$
 & $-46\pm27$ &--      & -- &$0.45\sim 0.60$~\cite{Amsler:2008zzb}\\
$S_{11}(1650)$& $41\pm 13$   & $53\pm16$     &  --  &--           &
$24\pm 7$
   & $-15\pm21$ & --     & --& $3\sim 10$\%~\cite{Amsler:2008zzb}\\
$D_{13}(1520)$& $-32\pm 7$  & $-24\pm 9$  &$113\pm 23$   &$166\pm 5$
& $-40\pm 8$   & $-59\pm9$  &$-124\pm 26$  &$-139\pm11$
& $0.05\pm 0.02$\%~\cite{Tiator:1999gr,Chiang:2001as}\\
$D_{13}(1700)$&  $-12\pm 4$ & $-18\pm13$  & $24\pm 8$   & $-2\pm 24$
 & $-2\pm 1$    & $0\pm 50$  & $-3\pm 1$  &$-3\pm 44$ & $0.1\pm 0.06$~\cite{Batinic:1995kr}\\
$D_{15}(1675)$&  --   &   $19\pm 8$   &--     &   $15\pm 9$ &
$-40$\footnote{These helicity amplitudes of $D_{15}(1675)$ are
obtained with $b_\eta=0.01$. }
& $-43\pm12$ & $-56^{a} $ & $-58\pm 13$& $0.0\pm 0.01$~\cite{Amsler:2008zzb}\\
\hline
\end{tabular}
\end{table}
\end{center}
\end{widetext}

\section{Summary}\label{summ}

In this work, we have studied the $\eta$ photoproduction off the
quasi-free nucleons within a chiral quark model. Our main motivation
is to study the bump-like structure observed in the $\eta n$ process
around $W\simeq 1.68$ GeV. We have achieved good descriptions of the
differential cross sections, total cross sections, and polarized
beam asymmetries for both processes in the three-quark scenario for
the baryon resonances.

It is found that the constructive interference between
$S_{11}(1650)$ and $S_{11}(1535)$ is responsible to the bump-like
structure around $W\simeq 1.68$ GeV in the $\gamma n\rightarrow \eta
n$ process, while the destructive interference between
$S_{11}(1650)$ and $S_{11}(1535)$ produces the shallow dip around
$W\simeq 1.68$ GeV in the total cross section of $\gamma
p\rightarrow \eta p$. Such interferences can lead to a sharp peak in
the ratio of $\sigma_n/\sigma_p$ at $W\simeq 1.68$ GeV. We also find
that no new resonances are needed for both of the processes to
interpret the observations in a rather broad energy region above the
threshold. We note that the importance of the interference between
$S_{11}(1650)$ and $S_{11}(1535)$ are also found for the $\pi
p\rightarrow \eta n$ process in our previous
studies~\cite{Zhong:2007fx}.

Apart from $S_{11}(1535)$ and $S_{11}(1650)$, $D_{13}(1520)$ also
plays an important role in both of the reactions. It accounts for
the major deviations of the angular distributions of the cross
section from an $S$ wave, and also produces large beam asymmetries
near threshold. The $u$ channel background can not be neglected for
both of the processes. It enhances the cross sections obviously. Our
calculation favors a much larger contribution from $P_{13}(1720)$
around its threshold. Since it is within the second orbital
excitation multiplets, such an enhanced strength could imply the
breakdown of the SU(6)$\otimes$O(3) symmetry. Further elaborate
treatment for higher excited states should be considered. We also
find that around $W\simeq 1.7$ GeV, $D_{15}(1675)$ is crucial for
the angular distributions of the cross section for $\gamma
n\rightarrow \eta n$, while $D_{13}(1700)$ plays an important role
in $\gamma p\rightarrow \eta p$.

It should be emphasized that the configuration mixing effects in the
$S$ and $D$ wave resonances are important for describing $\gamma
p\rightarrow \eta p$ near threshold. We have estimated the mixing
angles $\theta_S$ and $\theta_D$, which are $\theta_S\simeq
26^\circ$ and $\theta_D\simeq 21^\circ$. These mixings are also
included in $\gamma n\to \eta n$, although some ambiguities would
arise from the EM couplings. A much detailed study of this issue
will be reported elsewhere.

In this work, we have extracted the helicity amplitudes for the $S$
and $D$-wave resonances $S_{11}(1535)$, $S_{11}(1650)$,
$D_{13}(1520)$, $D_{13}(1700)$ and $D_{15}(1675)$ from the
reactions, which are compatible with the values from PDG.


\section*{  Acknowledgements }

This work is supported, in part, by the National Natural Science
Foundation of China (Grants 10775145, 11075051, 11035006), Chinese
Academy of Sciences (KJCX2-EW-N01), Program for Changjiang Scholars
and Innovative Research Team in University (PCSIRT, No. IRT0964),
Ministry of Science and Technology of China (2009CB825200), the
Program Excellent Talent Hunan Normal University, and the Hunan
Provincial Natural Science Foundation (11JJ7001).



\end{document}